\documentclass[5p]{elsarticle}
\usepackage{graphicx}
\usepackage[format=plain, font=small, labelfont=bf, justification = raggedright]{caption}
\usepackage{xspace}
\usepackage{amsmath}
\usepackage{amsfonts}
\usepackage{amssymb}
\usepackage{bm}
\usepackage{fancyhdr}
\usepackage{overpic}
\usepackage{xcolor}
\usepackage{algorithm}
\usepackage{algorithmic}
\usepackage{cuted}

\usepackage{hyperref} 
\hypersetup{colorlinks = true,linkcolor = blue, breaklinks = true}

\newcommand{\BVec}{\textbf{B}}    
\newcommand{\dd}{\text{d}}        
\newcommand{\phiNorm}{\widetilde{\varphi}} 
\newcommand{\IFunc}{\mathcal{I}}  
\newcommand{\CFunc}{\mathcal{C}}  
\newcommand{\SFunc}{\mathcal{S}}  
\newcommand{\dawson}{\mathcal{D}}  

\DeclareMathOperator{\csch}{csch} 
\DeclareMathOperator{\erfc}{erfc} 

\newcommand{\App}[1]{\ref{#1}}

\usepackage{subcaption}
\newcommand{\labelphantom}[1]{%
{\phantomsubcaption%
\label{#1}}%
}%
\makeatletter
\renewcommand\p@subfigure{\thefigure-}
\makeatother

\journal{Fusion Engineering and Design}

\begin{document}
\setlength{\parskip}{0pt}
\setlength{\belowcaptionskip}{0pt}


\title{Application of the Scrape-Off-Layer Fast Ion (SOLFI) code to assess particle motion in mirrors and tokamaks}

\author[PU,PPPL,TE]{X. Zhang}
\ead{laura.zhang@tokamakenergy.com}
\author[PU,PPPL,UO]{N. A. Lopez}
\author[CU]{A. O. Nelson} 
\author[CU]{L. Rondini} 
\author[PPPL,ITER]{F. M. Poli}
\address[PU]{Princeton University, Princeton, New Jersey 08540, USA}
\address[CU]{Columbia University, New York City, New York 10027, USA}
\address[PPPL]{Princeton Plasma Physics Laboratory, Princeton, New Jersey, 08540, USA}
\address[TE]{Author's current address: Tokamak Energy Ltd, Milton, Abingdon, United Kingdom}
\address[UO]{Author's current address: University of Oxford, Oxford, United Kingdom}
\address[ITER]{Author's current address:  ITER Organization, St Paul Lez Durance Cedex, France}

\begin{abstract}

This paper introduces the Scrape-Off-Layer Fast Ion (SOLFI) code, which is a new and versatile full-orbit Monte Carlo particle tracer developed to follow fast ion orbits inside and outside the separatrix in tokamaks. SOLFI is benchmarked in a simple straight mirror geometry, showing that the code conserves particle energy and magnetic moment, obtains the correct passing boundary for particles moving in the magnetic mirror field with an imposed electrostatic field, and correctly observes equal ion and electron current at the ambipolar potential predicted from analytical theory. This result has consequences for collisionless scrape-off-layers in spherical tokamaks. We then utilize SOLFI for fundamental physics studies in novel tokamak geometries, exploring the effect of shaping on the trapped particle fraction and bounce locations in tokamaks and demonstrating that negative triangularity can be used to maximize the fraction of particles bouncing in the good-curvature region, potentially leading to enhanced confinement. 
\end{abstract}

\maketitle

\pagestyle{fancy}
\lhead{Zhang, \textit{et al.}}
\rhead{SOLFI}
\thispagestyle{empty}

\section{Introduction}


Fast ions are produced in tokamak plasma through multiple ways, including neutral beam injection (NBI), ion cyclotron resonance heating (ICRH), and fusion reactions. These ions are born with energies much higher than the bulk thermal plasma and are then scattered and slowed down by the thermal population, depositing energy and momentum into the bulk plasma itself. This process is critical to the heating and sustained burning of the fusion plasma. However, these energetic particles are often also the source of trouble in varied ways. They are a source of free energy that can destabilize a wide spectrum of magnetohydrodynamic (MHD) fluctuations \cite{garcia2013fast, garcia2019active}. The resonance between the MHD modes and the energetic particles also transport these fast particles, sometimes resulting in degradation of  confinement and damage to the plasma facing components \cite{garcia2019active, white1995toroidal}.

Mitigation methods of these unfavorable Alf\'enic events include, among others, the manipulation of fast ion distributions through interactions with waves in the ion cyclotron frequencies, injected electron cyclotron resonance heating waves, and external application of resonance magnetic perturbation (RMP) fields \cite{garcia2019active}. The influence of the Scrape-off-layer (SOL) plasma is then two fold. On one hand, waves injected through the antennae must pass through the SOL to reach the confined plasma, experiencing scattering and losses along the way. Similarly, the plasma response to RMP fields also include significant changes to the fields near the separatrix, both inside and outside \cite{van2019alfven, van2015fast}. On the other hand, since the Larmor and drift orbits of fast ions are large especially on the low field side, these particles frequently enter the SOL and are subsequently influenced by the SOL conditions. The combination and potential synergy of these two factors suggest that including the SOL in fast ion modeling may lead to improved understanding of energetic particle transport and confinement and aid in future experimental design.

In this paper, we introduce the 3-D full-orbit Monte Carlo code, named the Scrape-Off-Layer Fast Ion (SOLFI) code, which is under development as a possible means to extend the modeling capabilities of the NBI code NUBEAM \cite{pankin2004tokamak} into the SOL. We benchmark the SOLFI code by simulating the particle losses in a collisionless magnetic mirror, which can be considered a simple approximation of a hot collisionless SOL in a low aspect ratio ST, such as the Lithium Tokamak eXperiment (LTX)~\cite{Majeski09} and the National Spherical Torus eXperiment Upgrade (NSTX-U)~\cite{Menard12}. We show the SOLFI code is able to obtain the correct trapped-passing boundary when a prescribed electrostatic potential is imposed within the mirror, and also correctly obtains zero net exit current when the electrostatic potential is set to be the ambipolar potential predicted from simple analytical theory. We then discuss the relevance of these results to tokamaks. Finally, we use the SOLFI code to study particle orbits in a tokamak geometry with varied triangularity. We find that negative triangularity maximizes the fraction of particles bouncing in the good-curvature region of the plasma volume, potentially leading to enhanced confinement. We also find that elongation has no impact on particle trapping behavior, contrary to what had been previously hypothesized.



This paper is organized as follows. In section~\ref{sec:code} we introduce the SOLFI code and discuss the key algorithms involved. In section~\ref{sec:mirrorVALIDATE} we present the benchmarking case of trapped particle orbits in a simple magnetic mirror with prescribed electrostatic potential. Section~\ref{sec:NT} then presents modeling results for particle orbits in a realistic tokamak geometry with negative triangularity. Finally, in section~\ref{sec:conclusion} we summarize our main results. Auxiliary calculations are presented in appendices. 




\section{The Scrape-Off-Layer Fast Ion (SOLFI) particle tracing code}
\label{sec:code}

SOLFI is a full-orbit Monte-Carlo particle tracer code that aims to simulate the fast ion orbits that traverse both the core and the SOL plasma in a tokamak. It was developed as a possible extension to the NBI code NUBEAM \cite{pankin2004tokamak} to include SOL effects and gain better understanding of how the SOL influences fast particle confinement. Though it is intended for use in tokamaks, the particle tracer code itself is flexible in terms of magnetic geometry (Cartesian or toroidal) and is capable of simulating the dynamics of non-relativistic charged particles of arbitrary species in a variety of magnetic and electric field configurations. In this section, we introduce the core numerical algorithms in the SOLFI code. 

First of all, the particle trajectories are obtained by integrating the appropriate equations of motion using the Boris algorithm \cite{boris1970relativistic}, which guarantees energy conservation \cite{qin2013boris}. Collisions between the tracer particles and the background plasma (assumed Maxwellian for simplicity), as described by the linearized collision operator, are also included by integrating the corresponding Langevin's equation using a combination of the stochastic Euler method and the ESEC method for pitch angle scattering \cite{zhang2020simulating, fu2020explicitly}. Similar to the Boris algorithm, the ESEC algorithm also conserves energy while also being a fast and explicit numerical scheme.

For collisions occurring within the SOL, a rapid search algorithm on an unstructured grid is required to obtain the local plasma parameters. A common solutions to this type of search problem is to use tree-based data structures, where a heavy construction cost up front lowers the computational cost of the search itself, such as the R-tree approach used the fast ion component of the integrated plasma simulation code JINTRAC \cite{romanelli2014jintrac}. Here we speed up the search even more by requiring that the SOL plasma background be represented by a Delauney triangulation obtained using the \verb|delaunator| library \cite{delaunator}. This allows neighboring triangles to be found quickly, such that the search can be done by the 'walking on a triangulation' algorithm listed in algorithm \ref{alg:triag}. With this algorithm, only the first query for a newly initialized particle takes more than 1 iteration to locate. All subsequent searches are done in time $O(1)$.

\begin{algorithm}[H]
	\caption{Walking on a Triangulation}
	\label{alg:triag}
	\algsetup{indent=2em}
	\begin{algorithmic}[1]
		\STATE Start with query point $P$ and initial guess $t$
		\WHILE{$P$ is not in triangle $t$}
		\STATE Find the centroid of the triangle $P_0$
		\STATE Connect $P$ and $P_0$. 
		\STATE Find the edge $e$ in $t$ that line segment $P-P0$ intersects with
		\STATE Find triangle $t'$ that shares $e$ with $t$
		\STATE Set current guess $t = t'$
		\ENDWHILE
	\end{algorithmic}
\end{algorithm}

\begin{figure*}
	\centering
	\includegraphics[width=0.32\textwidth]{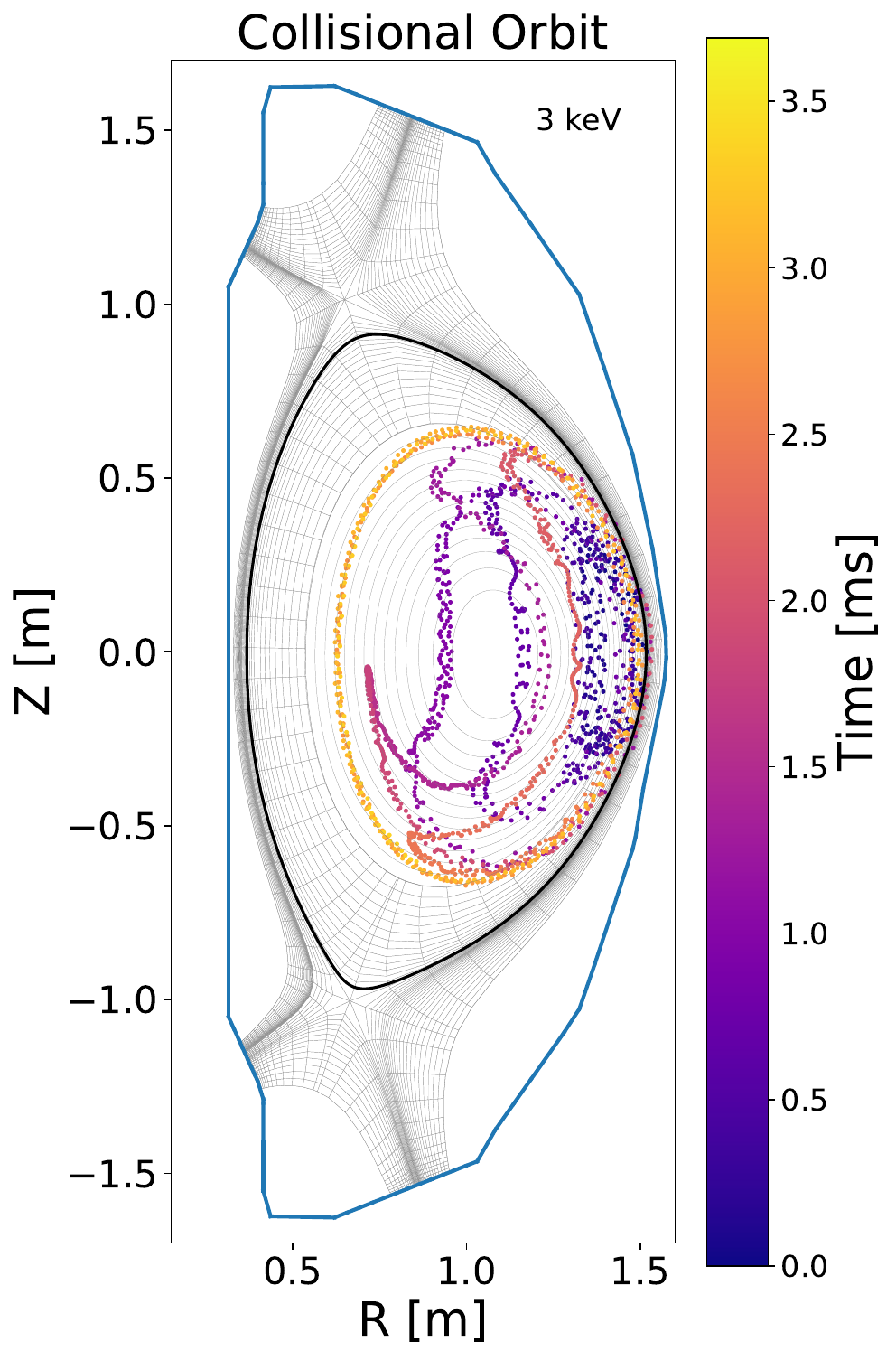}
	\includegraphics[width=0.32\textwidth]{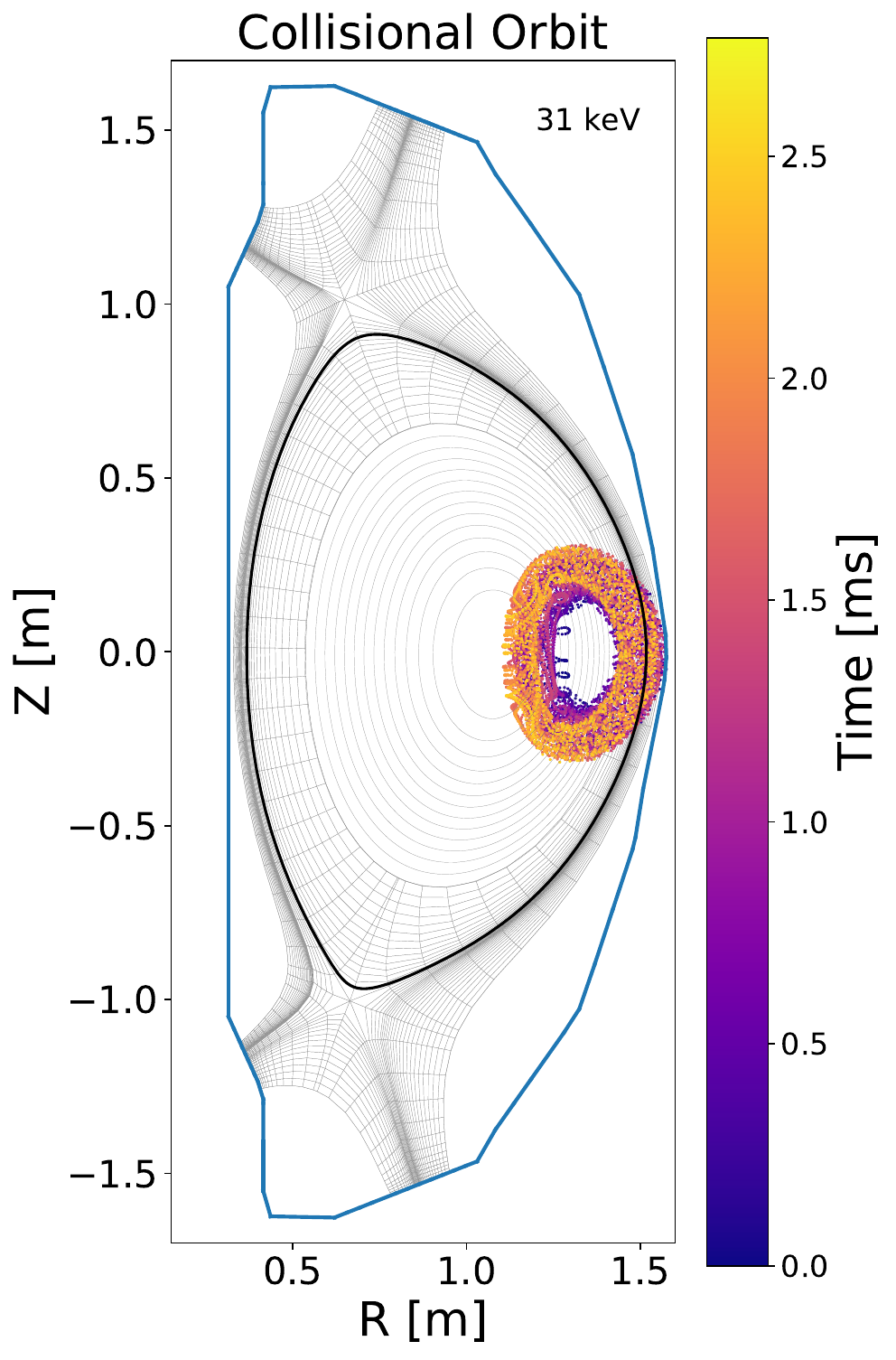}
	\includegraphics[width = 0.33\textwidth]{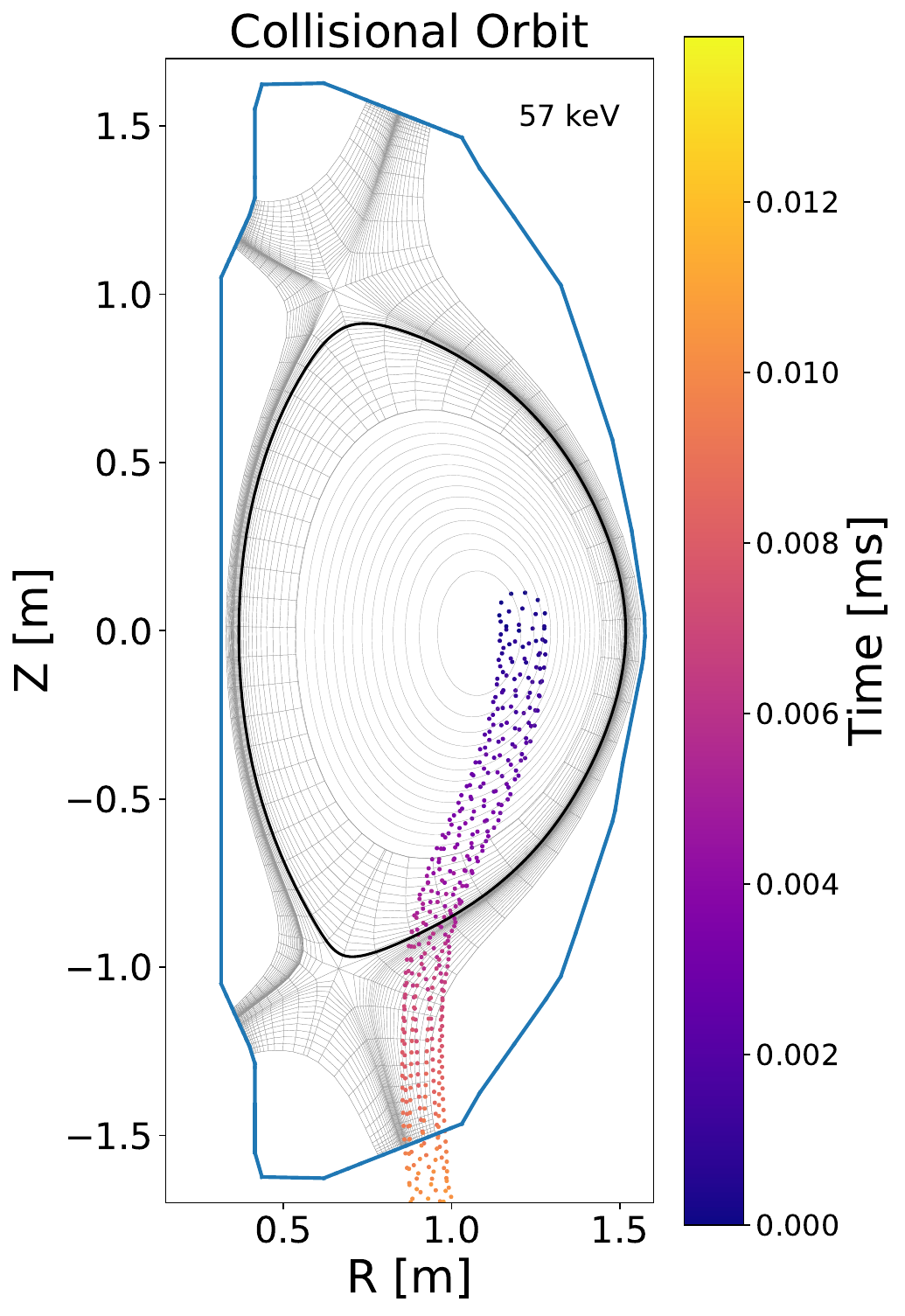}
	\caption{Example collisional particle orbits in an NSTX-U-like plasma for three different energies: $3\,$keV, $31\,$keV and $57\,$keV.}
	\label{fig:orbits_collisional}
\end{figure*}

The background electrostatic potential and magnetic field profiles can be initialized either analytically by the user, or inputted via the standard \verb|geqdsk| format for tokamaks. In the latter case, obtaining the field requires carefully differentiating the flux functions included in the input files. In SOLFI we choose to adopt a global rational barycentric interpolation of the stream functions. Barycentric interpolations are famously well-conditioned, and barycentric rational interpolation is proven to avoid Runge's phenomenon and to yield an interpolant that is infinitely smooth everywhere \cite{berrut14}. 

We therefore first construct a rational barycentric interpolant of the magnetic flux $\Psi(R_i, z_j)$ and poloidal current function $F(\psi_k)$ with respect to the values at the nodes $\{R_i, z_j\}$, where $i$ and $j$ range from 1 to $n_w$ and $n_h$ respectively. We then calculate the exact derivatives of the interpolant at the node values to obtain the magnetic fields via differentiation matrices as follows. Let $\Psi(R_i, Z)$ be the interpolating function at fixed radius $R = R_i$. Then in its barycentric form,
\begin{align}
	\Psi(R_i, z) = \sum_{j = 0}^{n_h}
    \left[ 
        \frac{\gamma_j \Psi_{ij}}{z - z_j} 
        \left( \sum_{k = 0}^{n_h} \frac{\gamma_k}{z - z_k} \right)^{-1}
    \right]
    ,
\end{align}

\noindent where $\Psi_{ij} = \Psi(R_i, z_j)$ and $\{\gamma_j \}$ are the interpolation weights (discussed below). Note that one necessarily has $\Psi(R_i, z) \to \Psi_{ij}$ when $z \to z_j$. Then, $\partial \Psi / \partial z$ at each grid point $(R_i, z_j)$ is given exactly as
\begin{equation}
	\frac{\partial \Psi}{\partial z}(R_i, z_j) = \sum_{l = 0}^{n_h} D_{ij} \Psi_{ij}
    ,
\end{equation}

\noindent where $D$ is the differentiation matrix
\begin{align}
	D_{jl} &=
	\begin{cases}
		\displaystyle\frac{\gamma_l}{\gamma_j} \frac{1}{z_j - z_l}       & \text{if  }  j \neq l, \\
		-\displaystyle\sum_{m = 0, m \neq j}^{n_h} D_{jm}        & \text{if  }  j = l.
	\end{cases}
\end{align}

Thus, the radial component of the magnetic field is
\begin{align}
	B_r(R_i, z_j) = 
    -\frac{1}{2\pi R} \frac{\partial \Psi}{\partial z}
    = -\frac{1}{2\pi R_i} 
    \sum_{l = 0}^{n_h} D_{jl} \Psi(R_i, z_l)
    .
\end{align}

\noindent Similarly, the $z$ component of the magnetic field is
\begin{align}
    B_z(R_i, z_j) = 
    \frac{1}{2\pi R}\frac{\partial \Psi}{\partial R}
    = \frac{1}{2\pi R_i} 
    \sum_{l = 0}^{n_w} \tilde{D}_{il} \Psi(R_l, z_j)
    ,
\end{align}

\noindent where $\tilde{D}$ is the radial differentiation matrix given by the analogous expression
\begin{align}
	\tilde{D}_{il} &=
	\begin{cases}
		\displaystyle\frac{\gamma_l}{\gamma_i} \frac{1}{R_i - R_l}        & \text{if  }  i \neq l, \\
		-\displaystyle\sum_{m = 0, m \neq i}^{n_w} \tilde{D}_{im}        & \text{if  } i = l.
	\end{cases}
\end{align}

\noindent Finally, the toroidal component of the magnetic field does not require differentiation, given simply by the barycentric formula
\begin{equation}
    B_{T} (R_i, z_j)
    =
    \sum_{k = 0}^{n_\Psi} 
    \left[
        \frac{\gamma_k F(\Psi_k)}{\Psi_{ij} - \Psi_k} 
        \left( R_i \displaystyle\sum_{k = 0}^{n_\Psi} \frac{\gamma_k}{\Psi_{ij} - \Psi_k}\right)^{-1}
    \right]
    ,
\end{equation}

\noindent where $n_\Psi$ is the number of grid points in the flux coordinate, and $\Psi_k$ is the corresponding value. Again, one has $B_T \to \Psi_k/R_i$ when $\Psi_{ij} \to \Psi_k$. For all cases, we choose the lowest-order weights $\gamma_k = (-1)^k$ both for the sake of simplicity and to avoid oscillations, since higher-order weights do not necessarily have decreasing Lebesgue constants (i.e., decreasing errors) \cite{berrut14}.

For the fastest evaluation, the resulting magnetic field can be interpolated bilinearly for evaluation at arbitrary locations. For smoother fields, bicubic spline can now be used to interpolate the magnetic field components directly. Both options are provided to the user. 


Example particle orbits in an NSTX-U-like plasma are shown in figure~\ref{fig:orbits_collisional}, demonstrating the wide variety of particle behavior that is captured by the model. Notably, the higher-energy orbits depicted here feature significant deviations into the SOL, highlighting SOLFI's ability to accurately simulate ion orbits that cross the tokamak separatrix. More will be said about particle orbits in tokamak geometry in Sect.~\ref{sec:NT}; first let us benchmark the code in a simpler mirror geometry.

\section{Validating SOLFI in a collisionless magnetic mirror}
\label{sec:mirrorVALIDATE}


\subsection{Context: collisionless scrape-off layers}
\label{sec:mirrorINTRO}

In recent years there has been an increased interest in the lithium conditioning of plasma-facing components (PFCs) as a means to better handle the divertor heat flux in reactor-grade tokamaks \cite{jaworski2013liquid, evtikhin2002lithium, nagayama2009liquid, rognlien2019simulations, poradzinski2019integrated, ono2020active, emdee2019simplified}. These lithium-coated PFCs (Li-PFCs) have also been shown to greatly improve the core confinement properties, to reduce oxygen impurity concentration, and to suppress damaging edge-localized modes \cite{majeski2010impact, mansfield2001observations, mirnov2003li, apicella2007first}. 

The use of Li-PFCs also has implications for the transport of thermal and fast ions because Li-PFCs reduce the edge neutral recycling, which in turn increases the temperature and decreases the density of the plasma in the SOL region \cite{kugel2008effect, Majeski17Compatibility, boyle2017observation, zakharov2004ignited, krasheninnikov2003lithium, zhang2019design, emdee2021predictive}. The neoclassical collisionality of the SOL is thus greatly reduced, and both thermal and fast ions can readily exhibit collisionless orbits that return to the core after crossing the separatrix into the SOL. It has been hypothesized that because of the low collisionality and the high mirror ratio of such SOLs, the particle confinement mechanism is different from those of a collisional, fluid-like SOL \cite{majeski2010impact, zhang2019design}, namely, the ions in the SOL should become mirror-trapped, with pitch-angle-scattering into the loss cone being the main loss mechanism. The formation of the ambipolar potential that maintains equal particle flux for ions and electrons is therefore different from standard sheath theory, which has consequences for the determining heat flux onto the target surfaces \cite{zhang2019design}. 

As a toy model for numerical validation of the SOLFI code, we attempt to loosely mimic the condition of the collisionless SOL by considering the particle dynamics in a straight magnetic mirror with prescribed electrostatic potentials. This problem is chosen specifically because it is simple enough to permit analytical solutions against which the code outputs can be compared.

\subsection{Simulation setup: Magnetic field and electric potential profiles of a simple magnetic mirror}
\label{sec:simpleMIRROR}

To perform the benchmarking, we require an analytical magnetic mirror field in 2-D slab geometry. Recall that if $\BVec(x,y)$ is a two-component divergence-free vector field, then it can be described by a scalar stream function $\zeta(x,y)$ such that $B_x(x,y) = \partial\zeta/\partial y$ and $B_y(x,y) = -\partial \zeta/ \partial x$. The level sets of $\zeta(x,y)$ then trace out the field lines of $\BVec(x,y)$. 

For an ideal magnetic mirror, a possible stream function is given by the 2-parameter family
\begin{equation}
    \zeta(x,y;R,\ell) = B_0 \, y\left[\frac{R-1}{\left( x/\ell \right)^4+1} + 1 \right]^{-1}
    ,
    \label{MirrorStream}
\end{equation}

\noindent where $R$ is the mirror ratio and $\ell$ roughly describes the width of the magnetic well. The magnetic field components are therefore given as
\begin{subequations}
    \begin{align}
        B_x(x,y;R,\ell) &= B_0 \frac{1+\left(x/\ell \right)^4}{R+\left( x/\ell \right)^4}
        , \\
        B_y(x,y;R,\ell) &=-B_0 \, (R-1)\frac{4 \left(x/\ell \right)^3 \left(y/\ell \right)}{\left[R+\left( x/\ell \right)^4 \right]^2}
        .
    \end{align}
    \label{mirrorFIELD}
\end{subequations}

\begin{figure}
    \centering
    \hspace{-10mm}\includegraphics[width=0.8\linewidth]{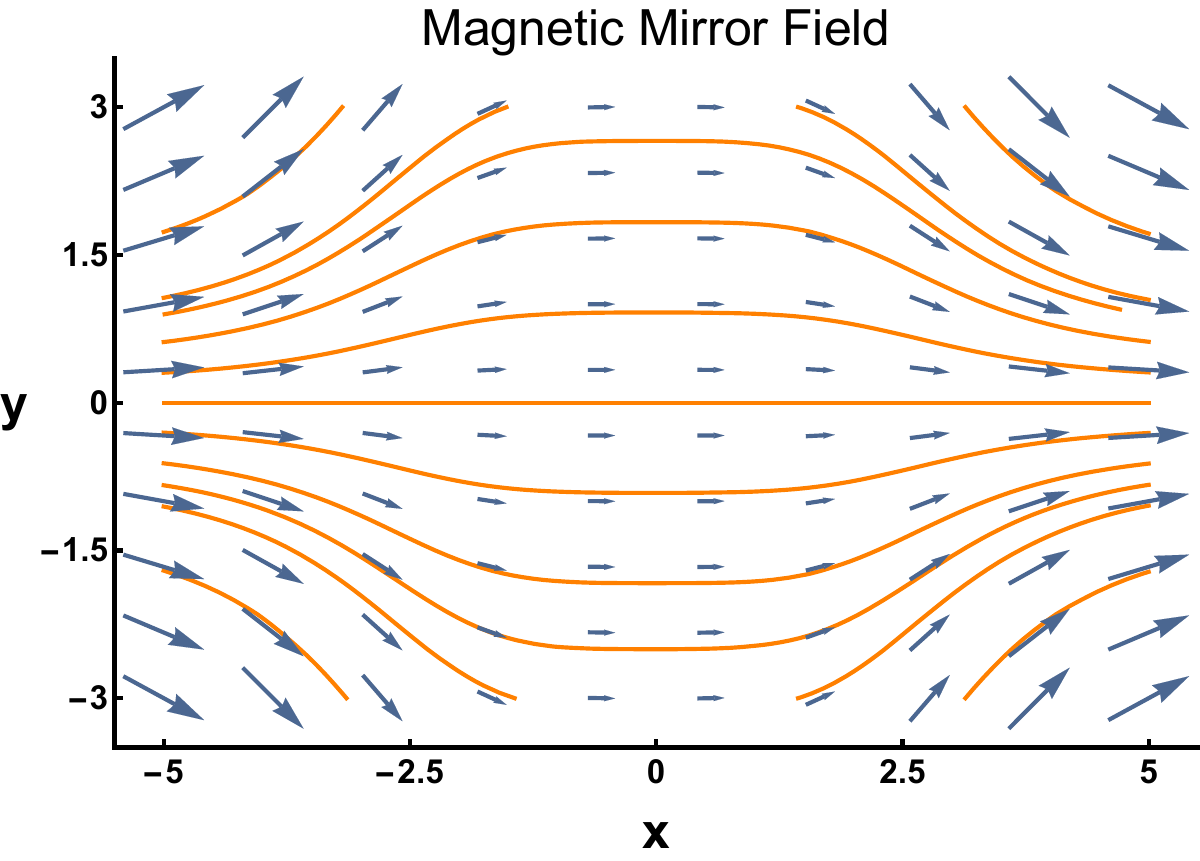}
    \caption{Magnetic mirror geometry used in the benchmarking of the SOLFI code. The level sets of the stream function $\psi(x,y)$ [Eq.~\ref{MirrorStream}] are shown in orange, while the strength and orientation of the resulting magnetic field [Eqs.~\ref{mirrorFIELD}] is depicted by the blue arrows. For this particular case, $R = 4$, $\ell = 3$, and $B_0 = 3$.}
    \label{Bgeometry}
\end{figure}

\noindent Note that along the mirror axis ($y=0$) one has as desired 
\begin{equation}
    |\BVec(\pm\infty,0;R,\ell)|/|\BVec(0,0;R,\ell)| = R
    .
\end{equation}

\noindent Sample field lines are shown in figure \ref{Bgeometry}. 

As for the imposed electric potential, it is sufficient for our purposes to choose the spatial profile
\begin{equation}
    \phi(x) = \phi_m\cos^6\left(\pi x/6\ell\right)
    ,
    \label{phiPROFILE}
\end{equation}

\noindent where $\phi_m$ is potential at midplane $x = 0$, and the length scale $\ell$ in Eq.~\ref{phiPROFILE} is the same $\ell$ that occurs in the magnetic field profile in Eq.~\ref{mirrorFIELD}. This specific functional form of $\phi$ is chosen to ensure that the collection plane $x = \pm 3\ell$ in our benchmarking study is accessible to all initialized particles, although we note that it is not necessarily the self-consistent potential field that would result from the particle motion. 

The simulation domain is chosen as $x \in [- 3\ell, 3\ell]$ and $y, z \in [-10\ell, 10\ell]$, where the magnetic axis is along the $\hat{x}$ direction. The large size of domain in the perpendicular direction is chosen to allow the perpendicular scale length of the magnetic field to be much larger than the particle Larmor radius, such that all cross-field drifts are negligible to facilitate comparisons with simple analytical theory. The choice to have the collection plane at $x = \pm 3\ell$ is simply because $|\BVec(3\ell,0; R, \ell)/ \BVec(0,0;R,\ell)|$ is satisfactorily close to the asymptotic mirror ratio $R$.


\subsection{Charged particle dynamics}


\begin{figure}
    \centering
    \includegraphics[width = 0.49\linewidth]{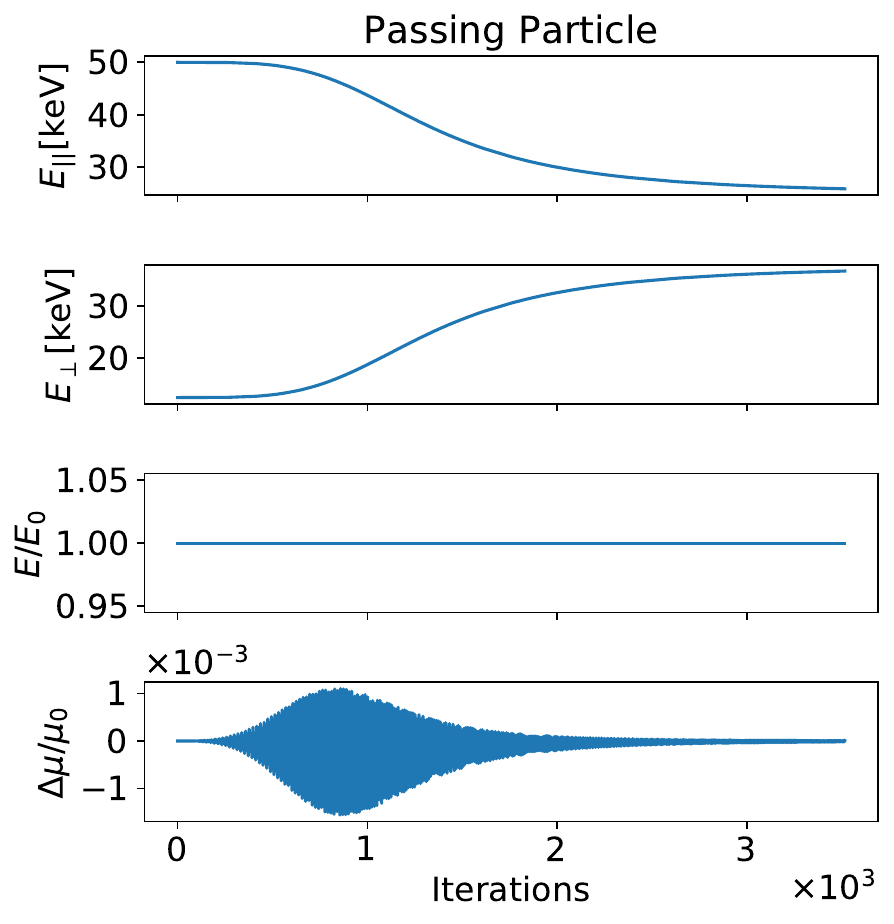}
    \includegraphics[width = 0.49\linewidth]{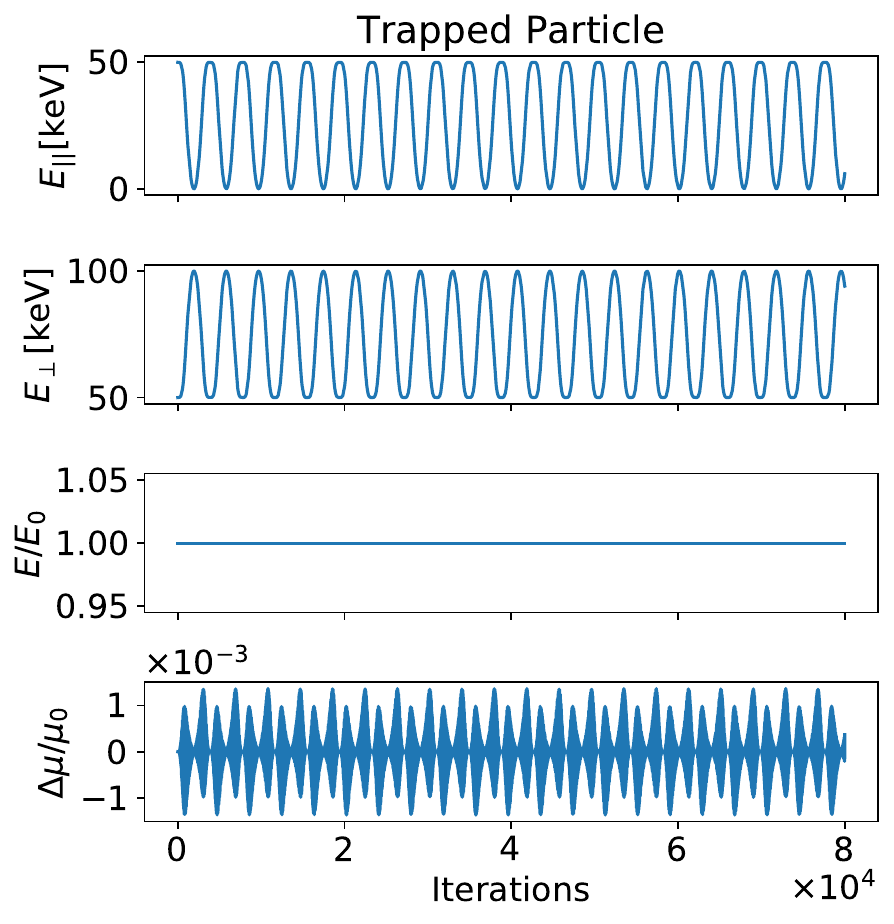}
    \caption{Time evolution of energy and magnetic moment of a passing particle (left) and a trapped particle (right) moving in the magnetic mirror field of Fig.~\ref{mirrorFIELD} with energies typical for NBI ions in NSTX-U. Only the first 20 bounces are shown for the trapped particle. Simulations for other energies are qualitatively similar.}
    \label{fig:EAndMu}
\end{figure}

\begin{figure*}
    \centering
    \includegraphics[width=0.35\linewidth]{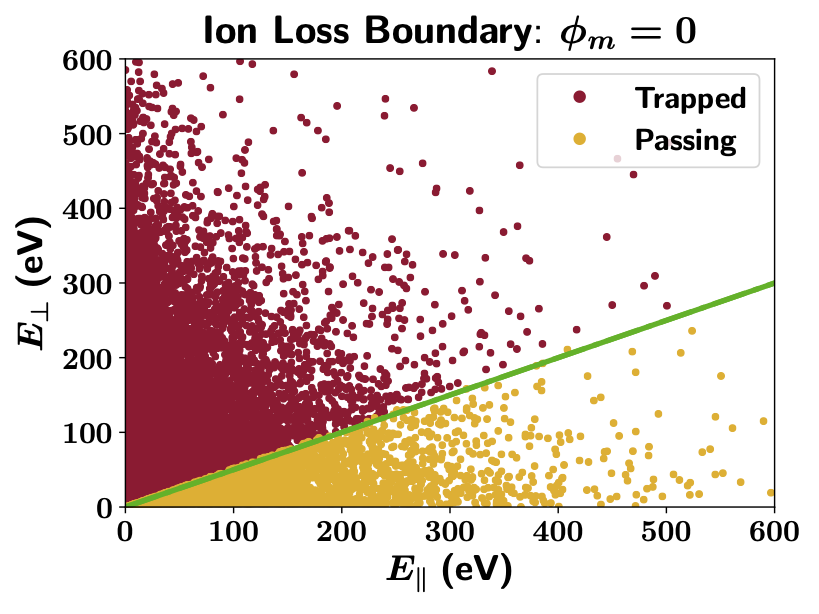}
    \hspace{10mm}\includegraphics[width=0.35\linewidth]{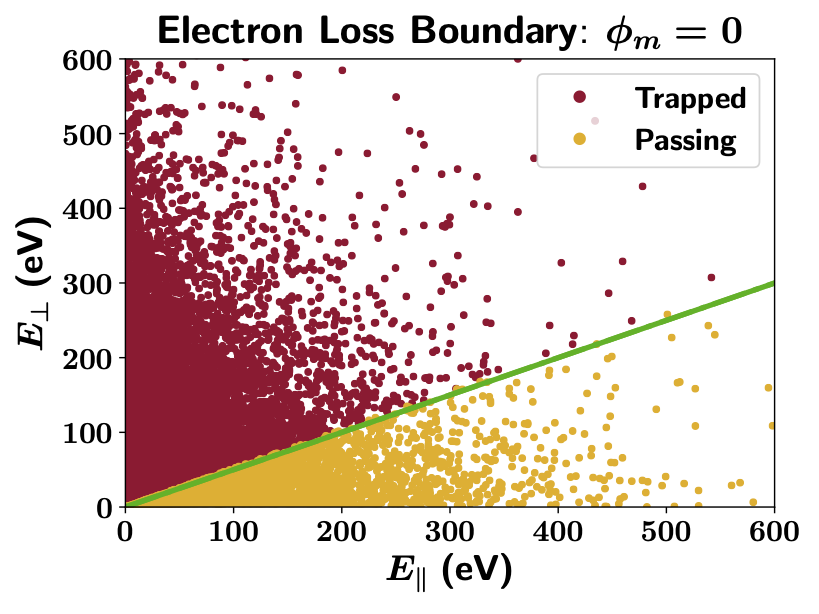}

    \includegraphics[width=0.35\linewidth]{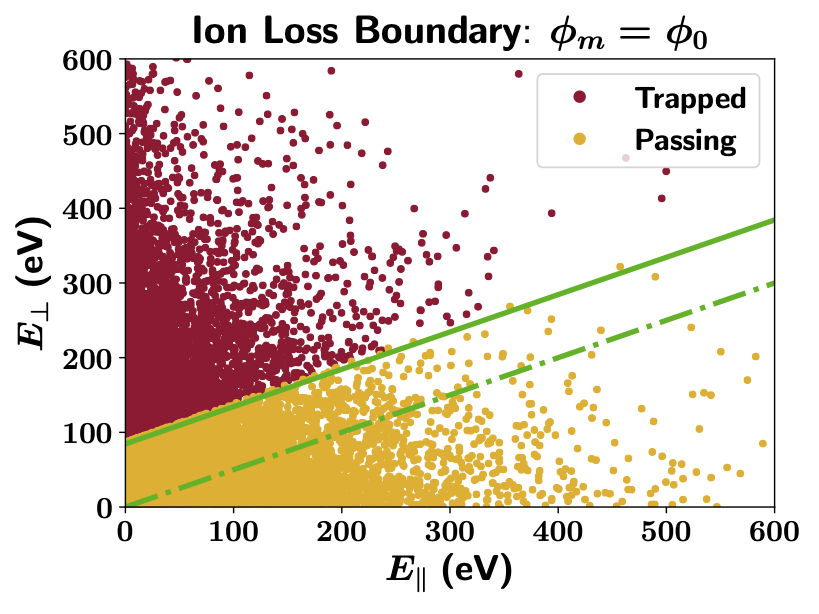}
    \hspace{10mm}\includegraphics[width=0.35\linewidth]{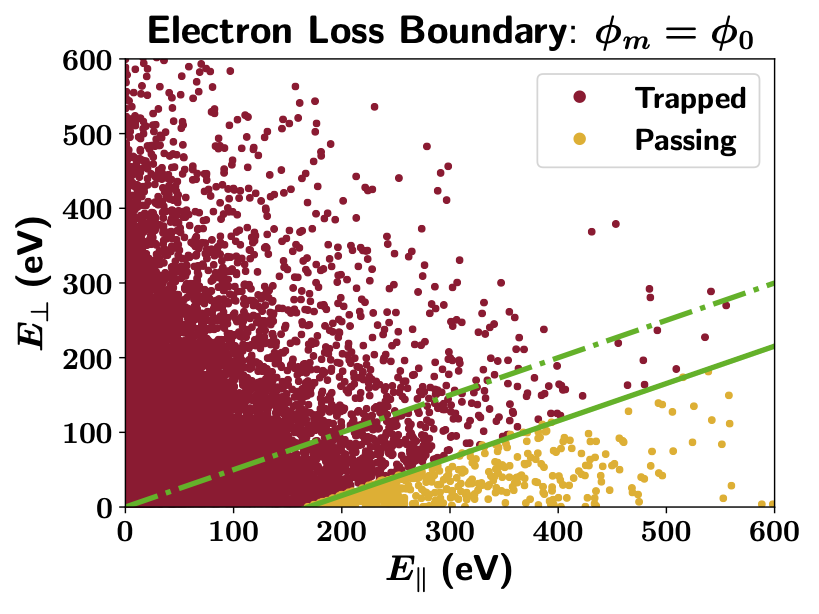}
    \caption{Ion and electron passing boundaries from full-orbit simulations. Each point represents the initial energy-space location of a particle. Ten thousand particles of each species are injected at the midplane and followed for up to 1 million time steps. Dashed green lines are the expected passing boundary in absence of electric potential $\phi_m$, whereas solid green lines are the expected loss boundaries with $\phi_m = \phi_0$ imposed, where $\phi_0$ satisfies Eq.~\ref{noCURRENT}.}
    \label{fig:lossCone}
\end{figure*}

The equations of motion for charged particles is simply the combination of Coulomb and Lorentz force, and is solved numerically using the Boris algorithm, as mentioned in the previous section. As is well-known (and reviewed in \App{app:mirror}), charged particles in a magnetic mirror with electrostatic potential exhibit either trapped or passing orbits along the mirror axis while conserving energy and magnetic moment. We confirm that the SOLFI code recovers this behavior in Fig. \ref{fig:EAndMu}, which shows the time evolution of parallel, perpendicular, and total energy as well as the magnetic moments for a representative passing and trapped particle with energies similar to those of NBI ions in NSTX-U. It can be seen that while the parallel and perpendicular energies transfer back and forth between bounces, the total energy of the particles are conserved exactly. The magnetic moment $E_\perp/B$ exhibits small, bounded oscillations around a constant value. 

To verify the passing-trapped boundary of sample particles is the same as that discussed in \App{app:mirror}, a collection of 20,000 particles are launched into the magnetic midplane $x = 0$. The particles are initialized according to a Maxwellian distribution with LTX-like thermal plasma parameters $T_i = T_e = 100$~eV. The mirror ratio is set to be $R = 3$. The particles are pushed up to 1 million timesteps, with $\sim 5$ steps per Larmor period. Particles that reach the mirror end at $x = \pm 3\ell$ are flagged as ``passing", or ``trapped" otherwise. Figure~\ref{fig:lossCone} shows the initial energy-space distribution of the test particles (corresponding to $t = 0$) and whether they are trapped or lost, with and without a prescribed potential drop. As seen from the figure, the simulated passing boundaries show excellent agreement with theoretical predictions. (Simulations were performed for a range of thermal plasma parameters, and the results are all qualitatively similar to those presented in figure~\ref{fig:lossCone}.)

\subsection{Ambipolar potential and ion confinement in a collisionless SOL}

As discussed in Sect.~\ref{sec:mirrorINTRO}, the ambipolar potential that arises in a mirror is fundamentally different from that which arises when a material surface (such as a probe) is placed in a non-magnetized plasma. In \App{app:ambipolar}, we develop and solve a simple analytical model to predict the ambipolar potential drop $\varphi$ that results in no net current exiting a magnetic mirror with given mirror ratio $R$. Here we verify this analytical calculation with the SOLFI code.

\begin{figure}
    \centering
    \includegraphics[width=0.9\linewidth, trim={1.2mm 0 0 0mm},clip]{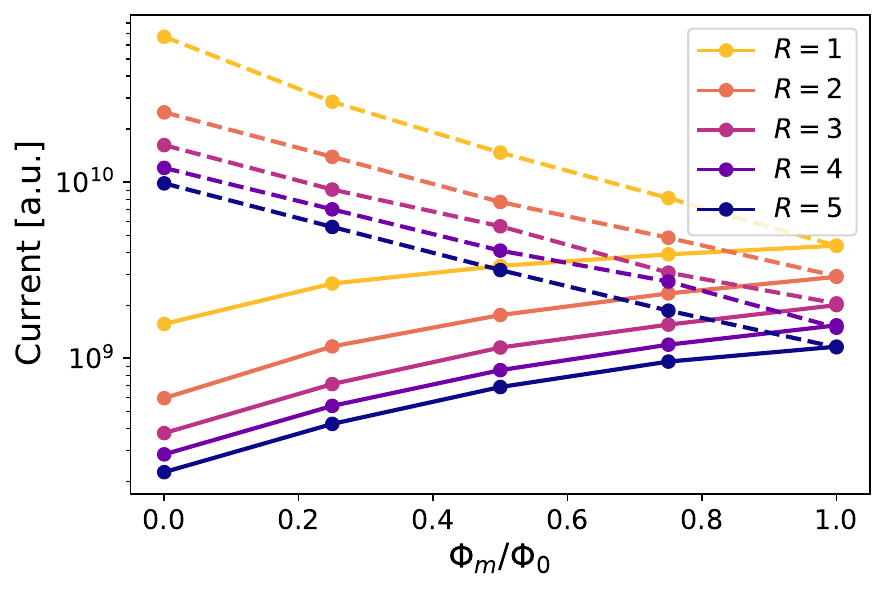}

    \includegraphics[width=0.9\linewidth]{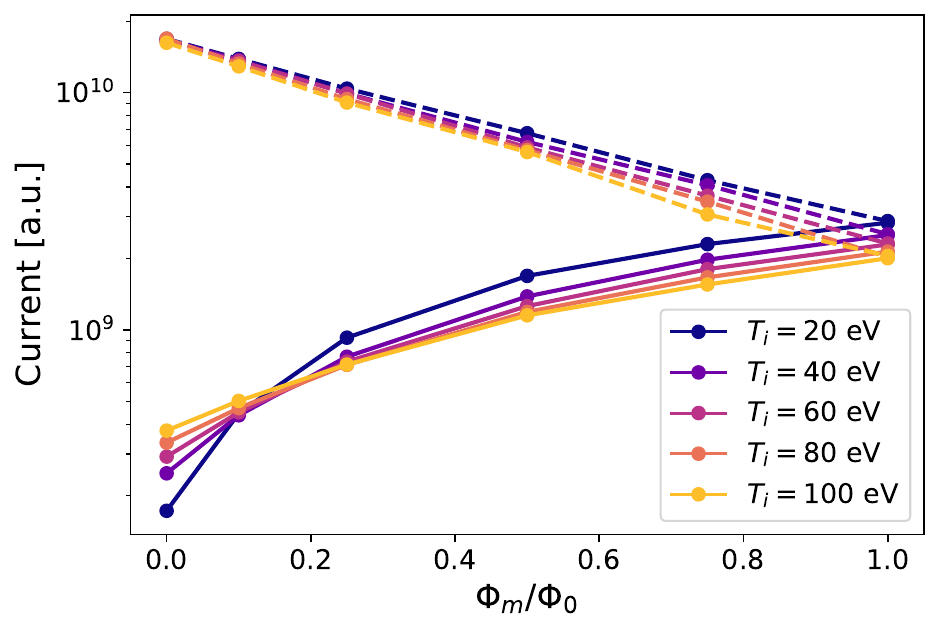}
    \caption{Convergence of electron (dashed lines) and ion (solid lines) passing current as the midplane potential $\phi_m$ approaches the theoretical value $\phi_0$ for the ambipolar potential that satisfies Eq.~\ref{noCURRENT}. (top) Mirror ratio $R$ is varied and $T_i = T_e$. (bottom) $T_i$ is varied, $T_e = 100$~eV and $R = 3$. Each data point results from the sum of $20,000$ particles.}
    \label{fig:currentScan}
\end{figure}

Specifically, let us confirm that the net passing current indeed vanishes when the prescribed potential is the solution of Eq.~\ref{noCURRENT}. To calculate the passing current, the parallel velocities of the particles are recorded when they reach the mirror end and then summed. Both electron and ion current are calculated as functions of the midplane potential $\phi_m$ with fixed electron temperature, but varying ion temperature and mirror ratio $R$. The resulting current is shown in figure~\ref{fig:currentScan}, where the ion and electron currents are seen to converge as the midplane potential approaches the predicted ambipolar value $\phi_0$ in all cases.

Although the details of the ambipolar potential $\phi_0$ that arises from differential transport in magnetic mirrors is reserved for \App{app:ambipolar}, we can understand the underlying physical dependencies of $\phi_0$ in relatively simple terms. The ambipolar potential depends on both the mirror ratio and the ratio between the ion and electron temperatures. When $R = 1$, i.e., when there is no magnetic mirror, $\phi_0$ decreases with increasing ion temperature, since the difference between the ion and electron thermal speeds decreases with increasing ion temperature. 

\begin{figure}
    \centering
    \includegraphics[width=\linewidth]{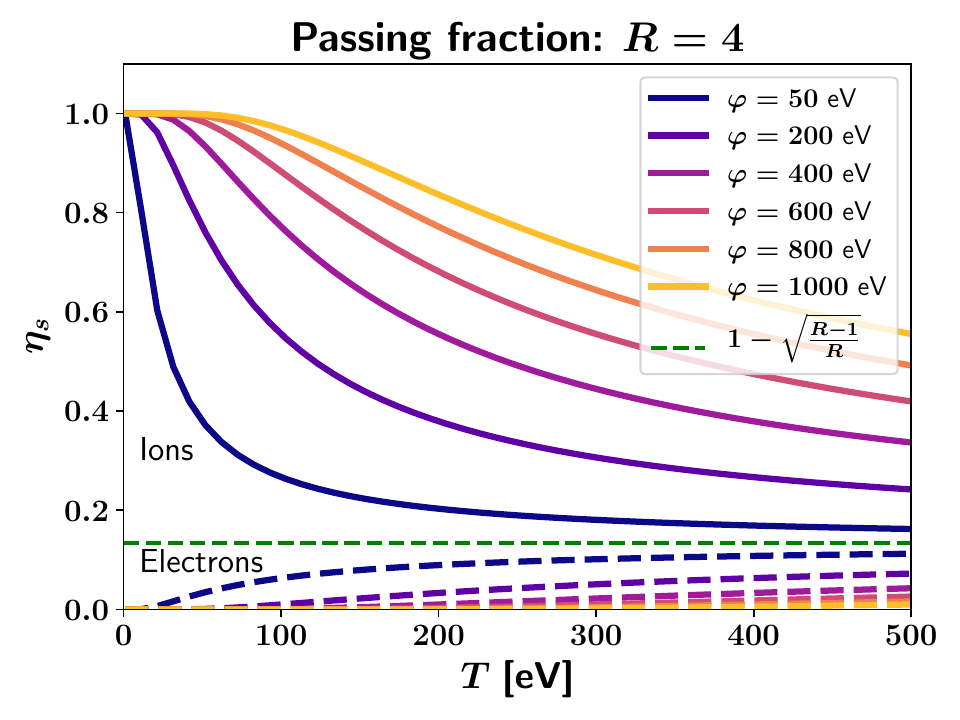}
    \caption{Passing-particle fraction $\eta_s$ for ions (solid lines) and electrons (dashed lines) in a magnetic mirror with $R = 4$ and various potential drops as the temperature is varied. The ion passing fraction is calculated with Eq.~\ref{etaION}, while the electron passing fraction is calculated with Eq.~\ref{etaELEC}.}
    \label{fig:passFRAC}
\end{figure}

This trend reverses for $R > 1$, however, where increasing the ion temperature increases $\phi_0$. This seemingly counter-intuitive result arises because the passing ion fraction $\eta_i$ decreases exponentially with increasing ion temperature, as shown in figure~\ref{fig:passFRAC}. (Details of the passing fraction calculation are included in \App{AppFRAC}.) Hence, the (now fewer) passing ions must be accelerated to higher velocities by a larger potential in order to balance the passing electron current. Lastly, at fixed ion and electron temperature, $\phi_0$ decreases with increasing $R$. This is simply because more electrons are mirror-trapped, resulting in a smaller electron passing current. 


Let us now discuss the implication of these results for collisionless SOLs in STs. Most notably, the exponential dependence of the ion-passing fraction on temperature has important consequences for ion-confinement estimates. Although the passing-particle fraction asymptotically approaches the oft-quoted neoclassical limit as $T/\varphi \to \infty$ (green dashed line in figure~\ref{fig:passFRAC}), this high temperature limit is rarely met in reality because $\varphi$ is typically comparable to $T_e$, being set by the ambipolar condition (Eq.~\ref{noCURRENT} in our case)~\cite{stangeby2000plasma}. Hence, the difference in trapped-particle fractions obtained via neoclassical estimate versus our result can be significant.

For example, the LTX SOL plasma typically has $T_e = 200$~eV, $T_i = 40$~eV, $R = 4$, and $\varphi = 0.5 \sim 3T_e = 100 \sim 600$~eV; we therefore estimate the ion-passing fraction as $\eta_i \gtrsim 0.8$. This is in drastic contradiction to the neoclassical estimate of $\eta_i \lesssim 0.2$ \cite{Majeski17Compatibility}, based on only considering the mirror ratio and not the plasma temperature as well. This trend also holds for a more general SOL plasma with $T_i \approx T_e$ and low collisionality: $\eta_i$ will be moderately lower than the above LTX-based estimate, but is still expected to be significantly higher than the neoclassical estimate due to the presence of the ambipolar potential. Hence, we conclude that despite the relatively high mirror ratio of the ST magnetic geometry, the ions in a collisionless SOL are still not mirror-confined, and their main loss mechanism remains direct streaming to the limiters.

\section{Impact of plasma shaping on trapped particle orbits}
\label{sec:NT}

\subsection{Context: Novel high-performance tokamak geometries}

\begin{figure*}
    \centering
    \labelphantom{fig:NTorbits-a}
    \labelphantom{fig:NTorbits-b}
    \labelphantom{fig:NTorbits-c}
    \labelphantom{fig:NTorbits-d}
    \labelphantom{fig:NTorbits-e}
    \labelphantom{fig:NTorbits-f}
    \includegraphics[width=0.7\linewidth]{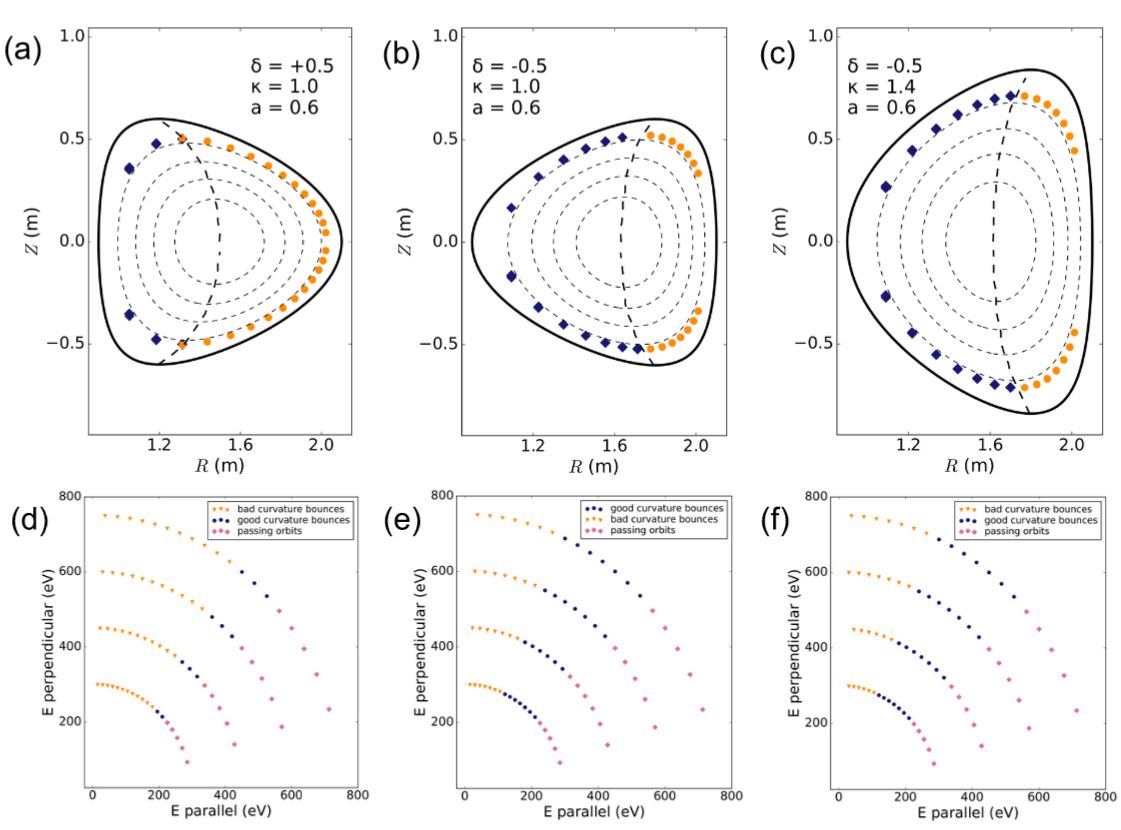}
    \caption{Bounce locations for edge-initiated particles of varying energy in a tokomak with (a) positive triangularity and (b), (c) negative triangularity. Yellow points indicate particles that bounce on the bad curvature side of the equilibrium (indicated by the dashed line) and blue points indicate particles that bounce on the good curvature side. In (d), (e) and (f), the energy distributions of particles in the equilibria shown in (a), (b) and (c), respectively, are shown.}
    \label{fig:NTorbits}
\end{figure*}

Plasma shaping is one of the most important considerations when designing and optimizing fusion reactors. While most modern tokamak research centers on an elongated ``D-shaped” plasma, emerging theoretical and experimental results have renewed interest in alternative shapes which could be advantageous for particle trapping, plasma confinement, power handling and/or MHD stability. For example, tokamak scenarios with negative triangularity (NT) have recently demonstrated improved core confinement while maintaining edge conditions that are favorable for power handling in fusion power plant scenarios \cite{Camenen2007, pochelon_recent_2012, Austin2019, coda_enhanced_2022, nelson_prospects_2022, nelson_robust_2023}. Work to understand the fundamental physics for this improved confinement is still ongoing, although heuristic explanations typically note (at least as a partial explanation) that trapped particles in NT spend more time in the good-curvature region, in turn reducing electron thermal transport by suppressing the trapped electron mode (TEM) instability \cite{kadomtsev_trapped_1971, Kikuchi, marinoni_diverted_2021, mackenbach_available_2023}. Even more exotic shapes, such as the ``comet" cross-section, which combines NT with oblate cross-sections characterized by having elongations $\kappa<1$, have been theorized to maximize the fraction of particles bouncing in the good curvature region, potentially leading to reactors with improved confinement and lower plasmas currents \cite{kesner_comet_1995}. 

In this section, we use the SOLFI code to investigate trapped orbits near the edge region of tokamaks, manipulating the triangularity and elongation of the plasma cross-section to better understand the effect of plasma shaping on trapped and passing orbits in tokamaks.

\subsection{Effect of triangularity}

We considered a large array of different equilibria with a variety of shaping parameters governing their magnetic geometry, and found that the most influential factor on the trapped particle dynamics is the plasma triangularity. The plasma triangularity $\delta$ is defined as $\delta \equiv (2 R_\mathrm{geo}-R_\mathrm{U} - R_\mathrm{L})/2a$, where $R_\mathrm{geo}$ is the geometric major radius, $R_\mathrm{U,L}$ are the major radii of the uppermost and lowest points along the plasma separatrix, respectively, and $a$ is the minor radius of the plasma. Notably, when $\delta<0$, the plasma X-points are pushed to the outboard side of the device with respect to the magnetic axis. This drags the separation between the so-called ``good" and ``bad" curvature regions, which are respectively stabilizing and destabilizing for MHD activity and turbulence \cite{Kikuchi}, outwards as well. In general, this allows the trajectory of a trapped particle to enter the good curvature region at lower energy values in NT than in traditional positive triangularity (PT) scenarios. 

The effect is illustrated in figure~\ref{fig:NTorbits}, which compares the bounce locations for a collection of particles with energies varying from 300 - 750 eV orbiting in a tokamak with PT versus one with NT. Because NT configurations place more of the plasma volume at lower magnetic field farther from the toroidal center, a larger fraction of the particles launched in the NT configuration bounce in the good curvature region. This effect can also be calculated analytically, revealing that the fraction of particles confined to the bad curvature region is a monotonic function of $\delta$ for configurations with specified magnetic field, aspect ratio and particle mass \cite{Sauter2016, marinoni_diverted_2021}.

Figures~\ref{fig:NTorbits-d} -- \ref{fig:NTorbits-e} also reveal that, while the overall particle energy does not directly impact its bounce location, the fraction of the kinetic energy parallel to the equilibrium magnetic field does (\App{app:mirror}). As the parallel velocity of the particle increases, the particle transitions from bouncing in the bad curvature region to bouncing in the good curvature region to eventually becoming a passing orbit. As shown in figure~\ref{fig:NTorbits-e}, the bad-to-good curvature transition energy is lower for NT than for PT, while the trapped-to-passing transition energy does not depend strongly on the plasma triangularity. These effects are consistent with the notion that there are more particles trapped in good-curvature regions for NT equilibria, and are all well-characterized by the SOLFI code, enabling statistical studies of trapped particle dynamics over a large selection of equilibria shapes. 

\begin{figure}
    \centering
    \includegraphics[trim={0 0 0 0.1}, width=\linewidth]{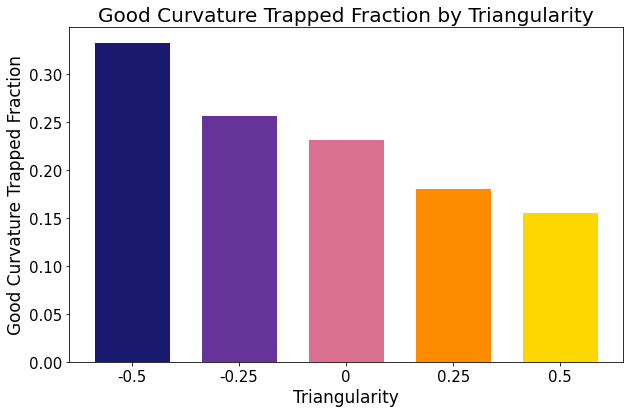}
    \caption{Percentage of trapped particles that bounce in the good curvature region as a function of the plasma triangularity.}
    \label{fig:NTgctf}
\end{figure} 

The direct influence of triangularity on the fraction of trapped particles that bounce in the good curvature region is further depicted in figure~\ref{fig:NTgctf}, which shows the results of an explicit scan in $\delta$ with  $\kappa$ and $a$ held fixed at the respective values of $\kappa= 1.0$ and $a = 0.6$. Clearly, as $\delta$ becomes increasingly negative, the good-curvature-trapped fraction increases. While many phenomena can impact the amplitude of turbulent transport caused by trapped-particle modes like the TEM, an enhancement of the good-curvature-trapped fraction is expected to lead directly to confinement improvements \cite{kadomtsev_trapped_1971, kesner_comet_1995}. This suggests that the NT geometry itself plays a vital role in reducing particle transport, in agreement with gyrokinetic observations of reduced electron thermal transport at strong NT \cite{Camenen2007, Austin2019, marinoni_diverted_2021}. Furthermore, since this effect is purely geometrical, the particle-trapping explanation is consistent with improved confinement observed both in ion-transport-dominated and electron-transport dominated regimes. Optimistically, this indicates that the improved core confinement observed in current NT experiments will be maintained in a demonstration device. However, additional gyrokinetic modeling and experimental validation are needed to move from this simple particle-orbit picture to a more robust explanation of NT behavior in burning plasma configurations.



\subsection{Effect of elongation}

Early analytical work suggested that the plasma elongation could also play a major role in determining particle orbits, potentially leading to performance gains in plasmas with oblate ``comet" cross-sections \cite{kesner_comet_1995}. Using SOLFI, the impact of elongation on particle trapping and on the good-curvature-trapped fraction were studied by launching a collection of particles in a tokamak geometry whose triangularity and minor radius were fixed at $\delta = -0.5$ and $a = 0.6$, respectively, while the elongation $\kappa$ was varied from $1.0$ - $1.6$. For illustration purposes, equilibria with two values of $\kappa$ are shown in figure~\ref{fig:NTorbits}. Elongation is observed to have little impact on the good-curvature-trapped fraction; however, further work using dedicated, high-fidelity tools is needed to assess to role of processional drifts (which may be reversed in shapes with low elongation) on plasma performance. 


\section{Conclusion}
\label{sec:conclusion}

In this paper we present the validation results of the full-orbit Monte Carlo particle code SOLFI in a simple magnetic mirror and in realistic tokamak geometry. For the mirror simulations, the passing-particle current and the ambipolar potential that balances electron and ion loss rates show excellent agreement with analytical theory. The conservation properties of energy and magnetic moment of the particle tracing algorithm is demonstrated, and as a result, the simulated passing/trapped boundary with and without an imposed electrostatic potential also show excellent agreement with predictions. Overall, the collisionless particle-tracing algorithm of SOLFI is consistent with expectations. 

We then consider the ambipolar potential $\varphi$ itself to gain understanding that might generalize from simple mirrors to collisionless SOLs. We show that the passing-particle fraction $\eta_s$ is exponentially modified by the potential; this means that thermal SOL ions in STs such as LTX remain untrapped by the magnetic-mirror field even when collisions are neglected. The exponential dependence of $\eta_s$ on $\varphi$ suggests that a strong electric field may significantly impact neoclassical particle confinement in the closed-field-line regions of tokamak plasmas as well. For example, strong radial electric fields are known to exist in the H-mode pedestal as a result of the pressure gradient \cite{wang2001neoclassical}. The electrostatic potential drop associated with this radial electric field across a trapped ion banana orbit is typically on the order of $T_i$~\cite{kagan2010enhancement}, which is not hot enough for the trapped-particle fraction to attain its asymptotic value that is often assumed when one attempts to estimate neoclassical processes based on trapped particles, such as the bootstrap current~\cite{kagan2010enhancement, watanabe1995effect}. 

Finally, we also used SOLFI to investigate the particle trapping in NT plasmas. We show that NT plasmas have a higher fraction of particles bouncing in the good-curvature region, and we also show that elongation does not have a significant impact. This geometric effect has been posited as a heuristic explanation for the improved confinement exhibited by NT plasmas \cite{marinoni_diverted_2021}, but a direct quantification of the effect has not previously been documented. That said, further studies are still needed to definitively link the increased good-curvature-trapped fraction exhibited by NT plasmas to the observed performance improvements. 


\section*{Acknowledgements}

The authors would like to acknowledge Elijah Kolmes and Ian Ochs for helpful conversation. X Zhang would like to thank Leonid Zakharov, Richard Majeski, and Nathaniel Fisch for helpful guidance. This material is based upon work supported by the U.S. Department of Energy, Office of Science, Office of Fusion Energy Sciences under Awards DE-AC02-09CH11466 and DE-SC0022272.

\section*{Data Availability}
The data that support the findings of this study are available from the corresponding author upon reasonable request.

\appendix

\section{Trapped-passing boundary for orbits in straight mirror with potential drop}
\label{app:mirror}

Here we review the behavior of particles in a magnetic mirror field with electrostatic potential for completeness.Consider a magnetic mirror field aligned along $\hat{x}$ in a two-dimensional ($2$-D) slab geometry with $\hat{z}$ an ignorable coordinate. Specifically, we take
\begin{equation}
    \BVec = B_x(x,y) \hat{x} + B_y(x,y) \hat{y}
\end{equation}

\noindent such that
\begin{subequations}
    \begin{equation}
        B_x(x,y) > 0
        , \quad
        B_y(x,0) = 0
    \end{equation}
    and for any $x_2 \ge x_1 \ge 0$,
    \begin{equation}
        |\BVec(x_2,y)| \ge |\BVec(x_1,y)|
        .
    \end{equation}
\end{subequations}

\noindent We assume $\BVec$ to be sufficiently strong such that, neglecting collisions, the particle transport is dominated by dynamics along $\BVec$. Let us also introduce a $1$-D electrostatic potential $\phi(x)$, which can either be externally imposed or be the ambipolar potential that arises from differential transport of electrons and ions. Finally, let us assume that the particle source is such that the distribution function is spatially confined near the mirror axis $y = 0$ with a fixed velocity profile at $x = 0$. Note that this initial condition implies that the particle dynamics are effectively $1$-D, with transverse drifts like the curvature drift and the $E \times B$ drift being negligible.

Suppose we are interested in the particle flux at some plane $x = L$. For a single particle of species $s$ initialized on the mirror axis $y = 0$, conservation of magnetic moment implies that \cite{Freidberg10}
\begin{equation}
    \frac{v^2_{\perp,L}}{|\BVec(L,0)|} = \frac{ v^2_{\perp,0}}{|\BVec(0,0)|}
    ,
    \label{MuCons}
\end{equation}

\noindent while conservation of energy implies that
\begin{equation}
    v^2_{\parallel,L} + v^2_{\perp,L} + \frac{2q_s}{m_s} \phi(L) = v^2_{\parallel,0} + v^2_{\perp,0} + \frac{2q_s}{m_s} \phi(0)
    .
    \label{ECons}
\end{equation}

\noindent Here, $m_s$ and $q_s$ are respectively the mass and charge of species $s$, while $v_{\perp,x}$ and $v_{\parallel,x}$ denote respectively the perpendicular and parallel components of the particle velocity with respect to the magnetic field at position $x$. 

Combining Eqs.~\ref{MuCons} and \ref{ECons} yields an expression for the exit velocity $v_{\parallel,L}$ in terms of the midplane velocities $v_{\parallel,0}$ and $v_{\perp,0}$ as
\begin{equation}
    v_{\parallel,L} = \sqrt{v^2_{\parallel,0} - (R-1)v^2_{\perp,0} + \frac{2q_s}{m_s}\varphi}
    ,
    \label{exitVEL}
\end{equation}

\noindent where we have defined the mirror ratio and the potential drop along the mirror axis as respectively
\begin{equation}
    R \doteq \frac{|\BVec(L,0)|}{|\BVec(0,0)|} \ge 1
    , \quad 
    \varphi \doteq \phi(0) - \phi(L)
    .
\end{equation}

\noindent The `passing region' is defined as the region of midplane velocities such that $v_{\parallel,L}$ is real, namely,
\begin{equation}
    v^2_{\parallel,0} - (R-1)v^2_{\perp,0} + \frac{2q_s}{m_s}\varphi \ge 0
    .
    \label{EQpassREG}
\end{equation}

\begin{figure}
    \centering
    \includegraphics[width = 0.9\linewidth,trim={0mm 0mm 0mm 0mm},clip]{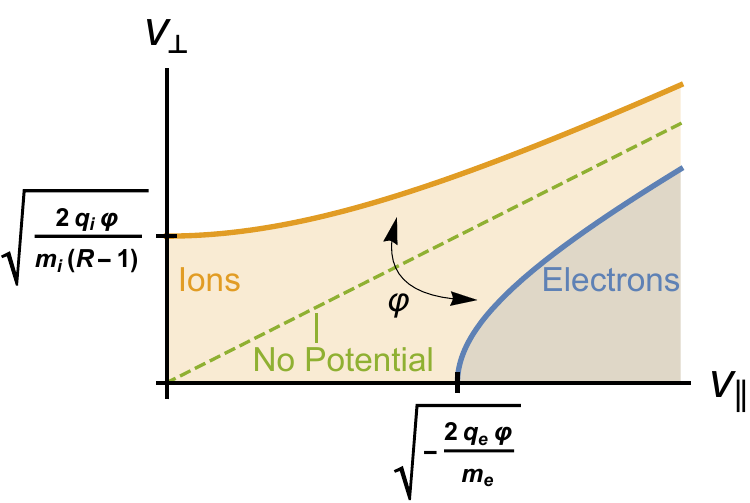}

    \hspace{5mm}\includegraphics[width = 0.78 \linewidth,trim={0mm 0mm 0mm 0mm},clip]{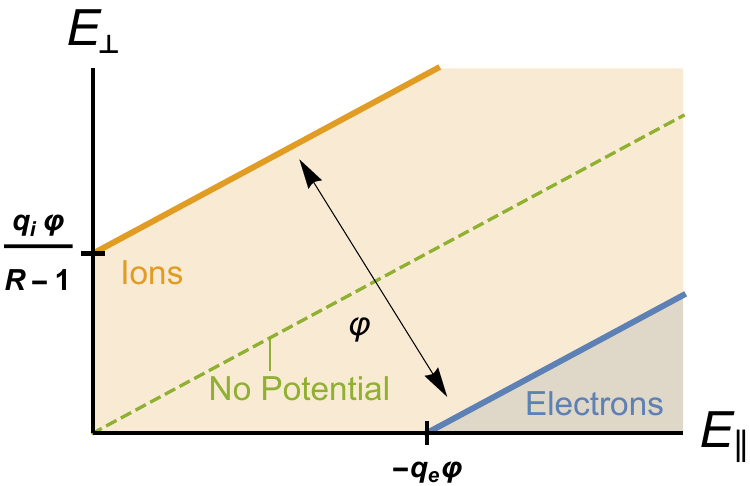}
    \caption{(Top) The passing region (shaded color) of midplane velocities $v_\perp$ and $v_\parallel$ for a magnetic mirror with mirror ratio $R$ and electrostatic potential drop $\varphi \ge 0$. (Bottom) Same, but in terms of midplane energies $E_\perp = m_s v_\perp^2/2$ and $E_\parallel = m_s v_\parallel^2/2$.}
    \label{passREG_v}
\end{figure}



\noindent As shown in figure~\ref{passREG_v}, this is a region bounded by a hyperbolic curve which either intersects the $v_{\perp,0}$ axis ($q_s \varphi > 0$), the $v_{\parallel,0}$ axis ($q_s \varphi < 0$), or becomes a straight line intersecting the origin ($q_s \varphi = 0$). The boundary becomes a straight line when written in terms of midplane energy, rather than midplane velocity. Note that for this work we require $x = L$ to be accessible, i.e., $v_{\parallel,x} \ge 0$ for all $0 \le x < L$ if $v_{\parallel,L} \ge 0$. This places constraints on the allowed spatial structure of $\phi(x)$.

\section{Loss current and ambipolar potential drop in simple magnetic mirror}
\label{app:ambipolar}

Given an initial midplane velocity distribution, the current outflow from a simple magnetic mirror can be obtained as the expectation value of Eq.~\ref{exitVEL} over the 'passing' region. Equating the current outflows of the positive and negative species will then give the ambipolar potential that will develop to maintain that no net current exists at the mirror exit wall. 

Let $f_s (v_{\parallel,0},v_{\perp,0} )$ be the midplane velocity distribution  for species $s$ (assumed to be gyrotropic for simplicity). Then, the collected current of species $s$ at the mirror exit $x = L$ is
\begin{align}
    \left\langle j_{s,L} \right\rangle 
    &= q_s \iint\limits_{\text{Passing}} \dd v_{\parallel,0} \, \dd v_{\perp,0} \,
    f_s\left(v_{\parallel,0},v_{\perp,0} \right)
    \nonumber\\[-3mm]
    &\hspace{15mm} \times
    \sqrt{v^2_{\parallel,0} - (R-1)v^2_{\perp,0} + \frac{2q_s}{m_s}\varphi}
    ,
    \label{collCUR}
\end{align}

\noindent where the integration is taken over the passing region (Fig.~\ref{passREG_v}). When the distribution function is a Maxwellian with density $n_s$ and temperature $T_s$, namely,
\begin{align}
    f_s\left(v_{\parallel,0},v_{\perp,0} \right) &= n_s v_{\perp,0}
    \sqrt{\frac{m_s^3}{2\pi T_s^3}}
    \nonumber\\
    &\hspace{4mm}\times
    \exp\left[
        -\frac{m_s}{2T_s}\left(v^2_{\parallel,0} + v^2_{\perp,0} \right)
    \right]
    ,
    \label{gyroMAXWELL}
\end{align}

\noindent then Eq.~\ref{collCUR} takes the form
\begin{equation}
    \left\langle j_{s,L} \right\rangle = q_s n_s \sqrt{\frac{8T_s}{\pi m_s}} \IFunc \left( \frac{q_s}{|q_e|} \frac{T_e}{T_s} \phiNorm;R\right)
    ,
\end{equation}

\noindent where $\phiNorm \doteq \frac{|q_e| \varphi}{T_e}$ is the potential drop normalized by the electron thermal energy and the (dimensionless) integral function $\IFunc(X;R)$ is given as
\begin{align}
	\IFunc( X ;R ) 
	&= \iint\limits_{\text{Passing}} \dd u \, \dd w \, 
	w \sqrt{u^2 - (R-1)w^2 + X}
	\nonumber\\[-4mm]
	&\hspace{22mm}\times
	\exp \left(-u^2 - w^2 \right)
	.
\end{align}

In terms of the nondimensionalized velocities $u$ and $w$, the passing region is defined as the region such that 
\begin{subequations}
    \begin{gather}
        \text{Max}\left[0, \sqrt{(R-1)w^2 - X}\right] \le u < \infty
        , \\
        0 \le w \le \sqrt{\frac{u^2 + X}{R-1}}
        .
    \end{gather}
\end{subequations}

\begin{figure}
    \centering
    \begin{overpic}[width=0.8\linewidth,trim={0mm 0mm 0mm 0mm},clip]{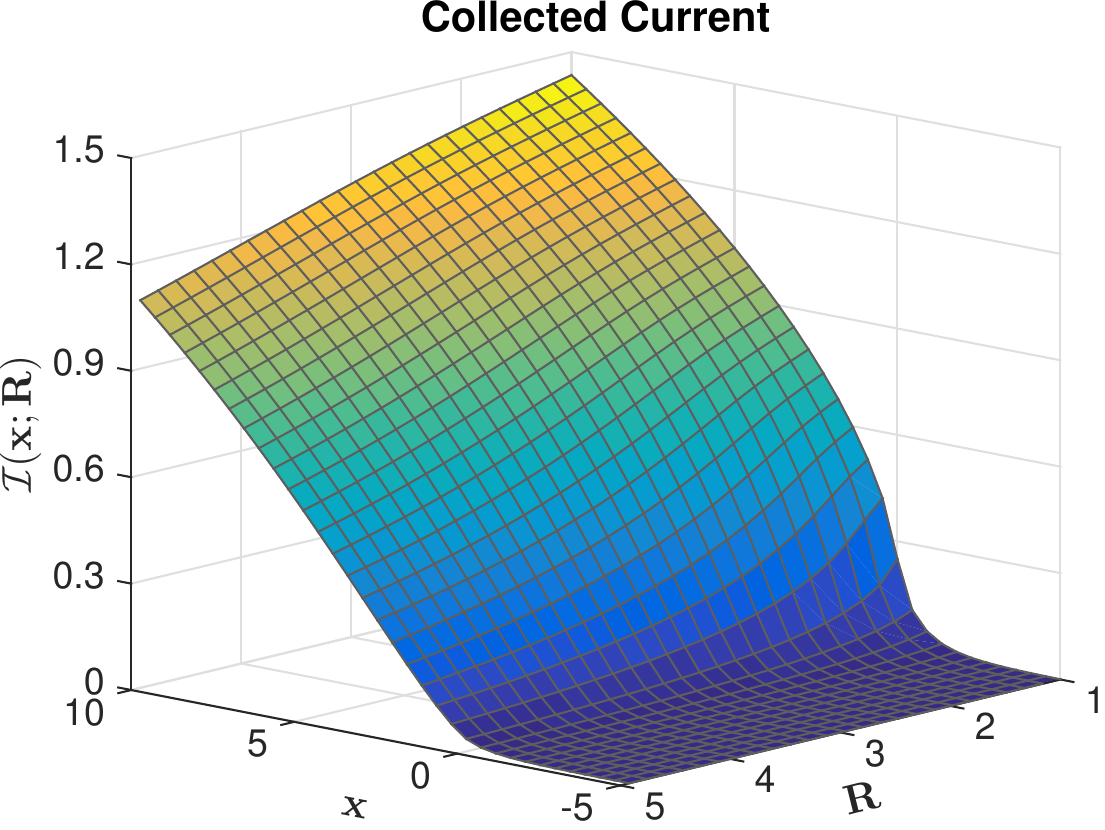}
        \put(15,15){\textbf{\Large(a)}}
    \end{overpic}
    
    \vspace{4mm}
    \begin{overpic}[width=0.8\linewidth,trim={0mm 0mm 0mm 0mm},clip]{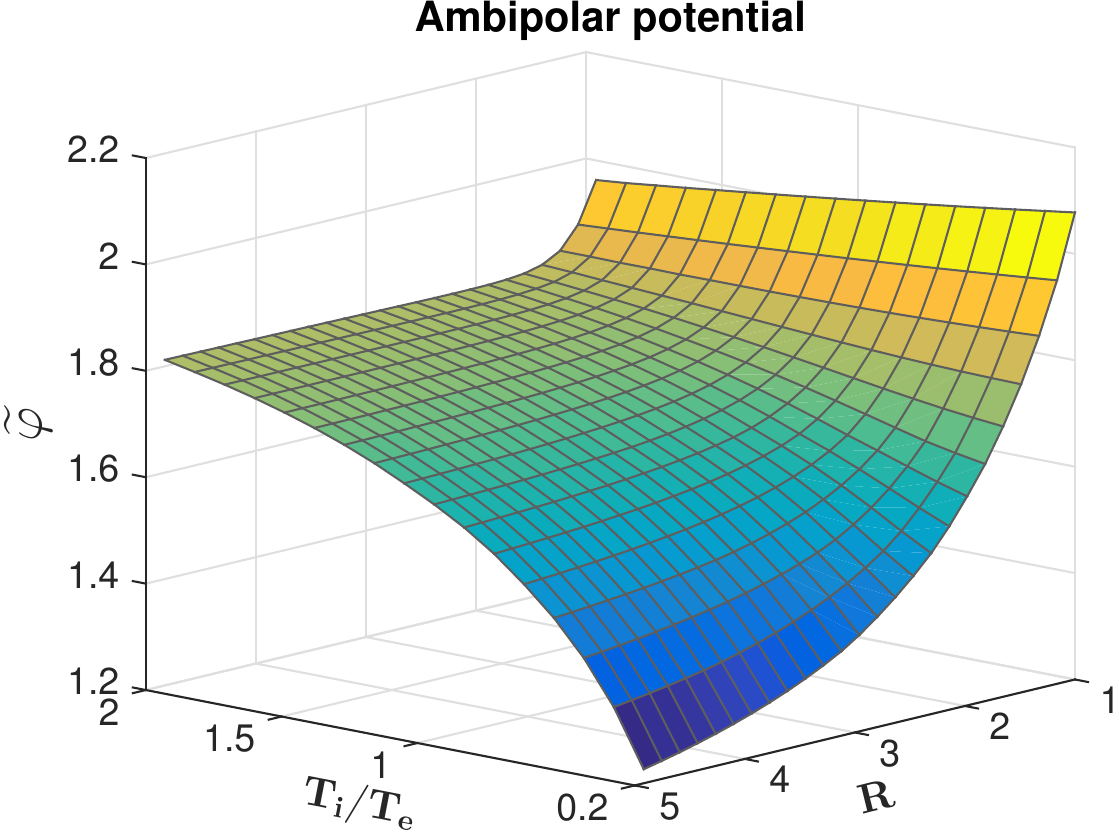}
        \put(15,15){\textbf{\Large(b)}}
    \end{overpic}
    \caption{\textbf{(a)} The integral function $\mathcal{I}(X;R)$, given explicitly in Eq.~\ref{IfuncSIMPLE}. \textbf{(b)} Ambipolar potential $\phiNorm$ obtained from solving Eq.~\ref{noCURRENT} over the depicted range of temperature ratio and mirror ratio $R$.}
    \label{surfaces}
\end{figure}

\noindent Hence, $\IFunc(X;R)$ can be written explicitly as
\newpage
\begin{strip}
\rule{\dimexpr(0.5\textwidth-0.5\columnsep-0.4pt)}{0.4pt}%
\begin{equation}
    \IFunc(X;R) = 
    \left\{
        \begin{array}{lr}
	        \int\limits_0^\infty \dd u \int\limits_0^{\sqrt{\frac{u^2 + X}{R-1}}} \dd w \, 
            w \sqrt{u^2 - (R-1)w^2 + X}
            \, \exp\left( -u^2 - w^2 \right)
	        & X \ge 0\\[5mm]
	        \int\limits_{\sqrt{|X|}}^\infty \dd u \int\limits_0^{\sqrt{\frac{u^2 - |X|}{R-1}}} \dd w \,
            w \sqrt{u^2 - (R-1)w^2 - |X|}
            \, \exp\left( -u^2 - w^2 \right)
	        & X < 0
	    \end{array} 
	\right.
	.
	\label{IfuncUNSIMPLE}
\end{equation}

\noindent Then, as shown in \App{AppINT}, $\IFunc(X;R)$ can be simplified as
\begin{equation}
    \IFunc(X;R) = 
    \left\{
        \begin{array}{lr}
	        \frac{X}{8}\left[K_1\left(\frac{X}{2} \right) + K_0\left(\frac{X}{2} \right) \right] \exp\left(\frac{X}{2}\right) 
	        -\frac{\sqrt{X(R-1)}}{4} \, \CFunc(X;R) \exp\left(\frac{X}{2}\right)
	        & X > 0\\[5mm]
	        \frac{1}{4} - \frac{R-1}{4\sqrt{R}}\, \csch^{-1}\left(\sqrt{R-1} \right) 
	        & X = 0\\[5mm]
	        \frac{|X|}{8}\left[K_1\left(\frac{|X|}{2} \right) - K_0\left(\frac{|X|}{2} \right) \right] \exp\left(-\frac{|X|}{2}\right)
	        - \frac{\sqrt{|X|(R-1)}}{4} \, \SFunc(|X|;R) \exp\left(-\frac{|X|}{2}\right)
	        & X < 0
	    \end{array} 
	\right.
	,
	\label{IfuncSIMPLE}
\end{equation}

\noindent where $K_\nu(X)$ is the modified Bessel function of the second kind~\cite{Olver10a}, and the functions $\CFunc$ and $\SFunc$ are given as
\begin{subequations}
    \begin{align}
        \label{cFUNC}
        \CFunc(X;R) &= \int_0^\infty \dd t ~\cosh\left(\frac{t}{2} \right) \dawson\left[\sqrt{\frac{X}{R-1}} \cosh\left( \frac{t}{2}\right) \right]
        \exp\left[-\frac{X}{2}\cosh(t)\right]
        , \\
        \label{sFUNC}
        \SFunc(X;R) &= \int_0^\infty \dd t ~\sinh\left(\frac{t}{2} \right) \dawson\left[\sqrt{\frac{X}{R-1}} \sinh\left( \frac{t}{2}\right) \right] \exp\left[-\frac{X}{2}\cosh(t)\right]
        ,
    \end{align}
\end{subequations}

\par
    \hfill
    \rule[0.5\baselineskip]{\dimexpr(0.5\textwidth-0.5\columnsep-1pt)}{0.4pt}
\end{strip}

\noindent with $\dawson(z)$ being the Dawson function~\cite{Olver10a,Press07}. In this form, $\IFunc(X;R)$ can be tabulated for discrete values of $X$ and $R$ using standard quadrature and special functions packages for subsequent interpolations. Figure~\ref{surfaces}(a) shows $\IFunc(X;R)$ for $X \in [-5, 10]$ and $R \in [1,5]$.

If the plasma is left to freely relax, then the potential drop $\phiNorm$ will be the ambipolar potential that ensures no net current is collected at the mirror exit $x = L$. For a two-component plasma, this means $\left\langle j_{i,L} \right\rangle + \left\langle j_{e,L} \right\rangle = 0$, or equivalently for specific case of Maxwellian midplane distributions,
\begin{equation}
    \sqrt{\frac{m_e}{m_i}\frac{T_i}{T_e}} \IFunc\left(\frac{q_i}{|q_e|}\frac{T_e}{T_i}\phiNorm;R \right)
    - \IFunc(-\phiNorm;R)
    = 0
    ,
    \label{noCURRENT}
\end{equation}

\noindent after using the quasineutrality condition $q_in_i = |q_e|n_e$. The resultant ambipolar potential $\phiNorm$ can thus be obtained from Eq.~\ref{noCURRENT} via standard root-finding methods. Figure~\ref{surfaces}(b) shows the ambipolar $\phiNorm$ for Hydrogen plasma with $T_i/T_e \in [0.2, 2]$ and $R \in [1,5]$.

\section{Derivation of Eq.~\ref{IfuncSIMPLE}}
\label{AppINT}

Here we derive Eq.~\ref{IfuncSIMPLE} from Eq.~\ref{IfuncUNSIMPLE}. Let us first consider the integration over $w$, which has the general form
\begin{equation}
	\int_0^{\sqrt{a}} \dd w
	\, w \sqrt{a-w^2}~e^{-w^2} 
	= e^{-a} \int_0^{\sqrt{a}} \dd t \, t^2 e^{t^2}
	,
\end{equation}

\noindent where we transformed the integration variable from $w$ to $t \doteq \sqrt{a - w^2}$. Note that $a = (u^2 + X)/(R-1)$ in terms of Eq.~\ref{IfuncUNSIMPLE}. Integration by parts then yields
\begin{align}
    e^{-a} \int_0^{\sqrt{a}} \dd t \, t^2 e^{t^2}
    &= \left. \frac{t}{2} e^{t^2 - a} \right|_0^{\sqrt{a}}
    - \frac{1}{2} e^{-a}\int_0^{\sqrt{a}} \dd t ~e^{t^2}
    \nonumber\\
    &= \frac{\sqrt{a}}{2} - \frac{1}{2}\dawson(\sqrt{a})
    ,
    \label{vPERPint}
\end{align}

\noindent where $\dawson(z)$ is the Dawson function~\cite{Olver10a,Press07}, defined as
\begin{equation}
	\dawson(z) = e^{-z^2} \int_0^z \dd t ~e^{t^2}
	.
\end{equation}

Let us first consider the case when $X > 0$. Substituting Eq.~\ref{vPERPint} into Eq.~\ref{IfuncUNSIMPLE} yields
\begin{align}
    \IFunc(X > 0) 
    &=
    \int\limits_0^\infty \dd u \, 
    \frac{\sqrt{u^2 + X}}{2} \,
    e^{-u^2}
    \nonumber\\
    &-
    \frac{\sqrt{R-1}}{2}
    \int\limits_0^\infty \dd u \, 
    \dawson\left( 
        \sqrt{\frac{u^2 + X}{R-1}}
    \right)
    e^{-u^2}
	.
\end{align}

\noindent The transformation $u = \sqrt{X} \sinh(t/2)$ then yields
\begin{align}
    \IFunc(X > 0) 
    &=
    \int_0^\infty \dd t \,
    \frac{X\left[\cosh(t) + 1 \right]}{8}
    \exp\left[\frac{X}{2}-\frac{X}{2}\cosh\left(t \right)\right] 
    \nonumber\\
    &\hspace{4mm}-\frac{\sqrt{X(R-1)}}{4} \, \CFunc(X;R) \exp\left(\frac{X}{2}\right)
	,
	\label{xPOSsimple}
\end{align}

\noindent where $\CFunc(X;R)$ is defined in Eq.~\ref{cFUNC}. By recognizing the Sommerfeld integral representation~\cite{Olver10a,Watson66}:
\begin{equation}
	K_\nu(z ) = \int_0^\infty \dd t ~\cosh\left(\nu t \right) \exp\left[-z\cosh (t )\right]
	,
	\label{besselDEF}
\end{equation}

\noindent then Eq.~\ref{xPOSsimple} can be simplified to the top line of Eq.~\ref{IfuncSIMPLE}.

Let us next consider the case when $X < 0$ and Eq.~\ref{IfuncUNSIMPLE} takes the form
\begin{align}
    \IFunc(X < 0) &= 
    \int\limits_{\sqrt{|X|}}^\infty \dd u \,
    \frac{\sqrt{u^2 - |X|}}{2} e^{-u^2}
    \nonumber\\
    &\hspace{-4mm}- \frac{\sqrt{R-1}}{2}
    \int\limits_{\sqrt{|X|}}^\infty \dd u \,
    \dawson\left( 
        \sqrt{\frac{u^2 - |X|}{R-1}}
    \right)
    e^{-u^2}
	.
\end{align}

\noindent The transformation $u = \sqrt{|X|} \cosh(t/2)$ then yields
\begin{align}
    \hspace{-2mm}\IFunc(X < 0) 
    &=
    \int_0^\infty \dd t \,
    \frac{X\left[1 - \cosh(t) \right]}{8}
    \exp\left[\frac{X}{2}+\frac{X}{2}\cosh\left(t \right)\right] 
    \nonumber\\
    &-\frac{\sqrt{X(1-R)}}{4} \, \SFunc(|X|;R) \exp\left( \frac{X}{2}\right)
	,
	\label{xNEGsimple}
\end{align}

\noindent where $\SFunc(X;R)$ is defined in Eq.~\ref{sFUNC}. As before, invoking Eq.~\ref{besselDEF} simplifies Eq.~\ref{xNEGsimple} to the bottom line of Eq.~\ref{IfuncSIMPLE}.

Finally, let us consider $X = 0$. After noting that
\begin{equation}
    \int_0^\infty \dd u \, 
    u \,
    e^{-u^2}
    = \frac{1}{2}
    ,
\end{equation}

\noindent substitution of Eq.~\ref{vPERPint} into Eq.~\ref{IfuncUNSIMPLE} yields
\begin{align}
    \IFunc(0) &= 
    \frac{1}{4}
    -
    \frac{\sqrt{R-1}}{2}
    \int\limits_0^\infty \dd u \, 
    \dawson\left( 
        \frac{u}{\sqrt{R-1}}
    \right)
    e^{-u^2}
    .
    \label{xZEROsimple}
\end{align}

\noindent By making use of the identity~\cite{Oldham09}
\begin{equation}
    \int_0^\infty \dd u \,
    \dawson(a u) e^{-b u^2}
    =
    \frac{\csch^{-1}\left( \frac{\sqrt{b}}{a} \right)}{2 \sqrt{a^2 + b}}
    , \quad
    a, \, b > 0
    ,
\end{equation}

\noindent one simplifies Eq.~\ref{xZEROsimple} to the middle line of Eq.~\ref{IfuncSIMPLE}.

\section{Passing-particle fraction}
\label{AppFRAC}

Here, we derive the passing-particle fraction of species $s$, denoted $\eta_s$, in an ideal mirror with potential drop $\varphi$ for the gyrotropic Maxwellian distribution function provided in Eq.~\ref{gyroMAXWELL}. By definition, when $q_s \varphi \ge 0$ one has
\begin{equation}
	\eta_s
	= 
	\frac{4}{\sqrt{\pi}}
	\int_{0}^\infty \dd u \, e^{-u^2} \int_0^{\sqrt{a}} \dd w \, w e^{- w^2}
	,
\end{equation}

\noindent or when $q_s \varphi < 0$ one has
\begin{equation}
    \eta_s
    =
    \frac{4}{\sqrt{\pi}}
    \int_{\sqrt{\frac{|q_s\varphi|}{T_s}}}^\infty \dd u \, e^{-u^2} \int_0^{\sqrt{a}} \dd w \, w e^{- w^2}
    ,
\end{equation}

\noindent where we have introduced $a \doteq (u^2 + q_s\varphi/T_s)/(R-1)$. We readily evaluate
\begin{equation}
    \int_0^{\sqrt{a}} \dd w \, w \, e^{- w^2}
    = \frac{1}{2} 
    - \frac{1}{2} e^{-a}
    .
\end{equation}

\noindent Hence, for $q_s \varphi \ge 0$ we obtain
\begin{align}
    \eta_s &= \frac{2}{\sqrt{\pi}}
    \left\{ 
        1 
        - \sqrt{\frac{R-1}{R}} 
        \exp \left[- \frac{q_s \varphi_s}{T_s (R-1)}\right]
    \right\}
    \nonumber\\
    &\hspace{4mm}\times
    \int_0^\infty \dd u\, \exp\left(-u^2\right)
    \nonumber\\
    &=
    1 
    - \sqrt{\frac{R-1}{R}} 
    \exp \left[- \frac{q_s \varphi_s}{T_s (R-1)}\right]
    .
    \label{etaION}
\end{align}

\noindent Note that when $\varphi = 0$, $\eta_s$ becomes independent of the species temperature $T_s$.

By making use of the identity
\begin{equation}
    \int_a^\infty \dd x \, e^{-bx^2} 
    = \frac{1}{2} \sqrt{\frac{\pi}{a}} \, \erfc\left(a\sqrt{b} \right) 
    , \quad
    a, \, b > 0
    ,
\end{equation}

\noindent one can show that for $q_s \varphi < 0$, $\eta_s$ takes the form
\begin{align}
    \eta_s
    &= 
    \text{erfc}\left(\sqrt{\frac{|q_s\varphi|}{T_s}} \right) 
    \nonumber\\
    &- \sqrt{\frac{R-1}{R}}
    \erfc\left(\sqrt{\frac{|q_s\varphi|}{T_s} \frac{R}{R-1}} \right) \exp\left[\frac{|q_s\varphi|}{T_s(R-1)}\right]
    .
    \label{etaELEC}
\end{align}

\noindent Representative traces of Eq.~\ref{etaION} and Eq.~\ref{etaELEC} are shown in figure~\ref{fig:passFRAC} of the main text. Clearly, the fraction of passing ions $\eta_i$ decreases exponentially with increasing temperature. In fact, the passing fraction of both species approach the asymptotic value $1 - \sqrt{(R-1)/R}$ as the temperature is increased.

\bibliography{Biblio.bib}
\end{document}